\documentstyle[aps,preprint,amssymb,floats,psfig]{
revtex}

\begin{document}
\draft
\tighten
\preprint{
        \parbox{1.5in}{%
           \noindent
           PSU/TH/190}
	}
\title{Probing small-$x$ parton densities in proton-
	   \\
	proton (-nucleus) collisions in the very forward direction}	
\author{
   Lyndon Alvero, John C. Collins, Mark Strikman, and
   J.J. Whitmore}
\address{Physics Department, 104 Davey Lab, Pennsylvania State University,
       University Park, PA 16802-6300,
      U.S.A.}
\date{27 October 1997}
\maketitle
\begin{abstract}
We present calculations of several $pp$ scattering cross sections
with potential applications at the LHC.
Significantly
large rates for momentum fraction, $x$, as low as $10^{-7}$ are
obtained, allowing for possible extraction of quark and gluon
densities in the proton and nuclei down to these small $x$ values
provided a detector with good acceptance at maximal rapidities is used.
\end{abstract}


\section{Introduction}

In this paper, we study 
the measurement of quark and gluon distribution
functions inside the proton at very small momentum fractions $x$.
We consider several processes in $pp$ collisions
at a center-of-mass energy $(\sqrt{s})$ of 14 TeV and show that the 
range of useful
measurements extends down to about $x \sim 10^{-6}$ for most
processes, and even down to $x \sim 10^{-7}$ for the Drell-Yan
process.~\footnote{Similar considerations are valid for the
Tevatron collider where due to the smaller $\sqrt{s}$ the relevant
regions in $x$ are scaled up by a factor of $\sim 60$.}
Such measurements require a detector which has
sufficient acceptance at maximal rapidities.
Plans for building
a detector of this type (FELIX) are under discussion at the LHC
\cite{lhcloi}.
We show how the parton densities may be determined.  The
event rates are high; to estimate them, we use
the CTEQ3M distributions \cite{cteq3m} to provide an
extrapolation from the region where measurements currently exist,
which is $x\geq 10^{-4}$.

Although our results are obtained for proton-proton interactions,
a similar analysis can easily be applied to proton-nucleus
interactions in the same accelerator.
For example, in the case of proton-calcium
collisions, at say a center-of-mass energy of 63 TeV, the
same region, down to $x \sim 10^{-6}$ and lower,
can also be probed.  We show
how data from such studies would provide information on the
parton densities in nuclei at both large and small $x$.  Previous
data for hard processes on nuclei have been confined to
fixed-target energies, and so the range of processes for which
perturbative QCD (pQCD) calculations are reliable is limited.

At the LHC, the integrated luminosity for a proton-proton luminosity of
$10^{31}\,{\rm cm}^{-2}\,{\rm s}^{-1}$ and a run-time of
$10^{7} \, {\rm sec}/{\rm yr}$ is $100 \, {\rm pb}^{-1}$.
However, there is a loss of luminosity in proton-nucleus collisions;
for protons on calcium, a luminosity of 
$10^{30}\,{\rm cm}^{-2}\,{\rm s}^{-1}$
with a shorter running time is envisaged.
This loss of luminosity
is to a large extent compensated by increased cross
sections, which are approximately proportional to the mass number
$A$ for large $x$.  At small $x$, the cross sections presumably
behave more like $A^{2/3}$, but as we will see, the event rates are
so large that the resulting loss of event-rate relative to
proton-proton collisions will still enable a lot of physics to be
probed.

\section{Kinematics and cross sections}

We consider the following hadronic processes
at center of mass energy $\sqrt {s} = 14\ {\rm TeV}$ with $s=(p_1+p_2)^2$
\begin{eqnarray}
p(p_{1})+p(p_{2})&\to & {\rm jet} + \gamma + \ 
{\rm X} \ \ \ \ \ \ \ \ ({\rm J}\gamma), \nonumber \\
p(p_{1})+p(p_{2})&\to & l{\bar l} + \ 
{\rm X} \ \ \ \ \ \ \ \ \ \ \ \ \ \ ({\rm DY}), \nonumber \\
p(p_{1})+p(p_{2})&\to & {\rm jet}_1 + {\rm jet}_2 + \ 
{\rm X} \ \ \ \ ({\rm JJ}), \nonumber \\
p(p_{1})+p(p_{2})&\to & {\rm Q} + {\bar {\rm Q}} + \ 
{\rm X} \ \ \ \ \ \ \ \ ({\rm HQ}), \nonumber \\
p(p_{1})+p(p_{2})&\to & W/Z + \ {\rm X} \ \ \ \ \ \ \ \ \ \ ({\rm VB}), 
\label{gnricpro}
\end{eqnarray}
where 
$\gamma$ denotes a photon, $l{\bar l}$ a lepton pair,
${\rm Q}({\bar {\rm Q}})$ a heavy quark (antiquark) and
$W,Z$ are the weak vector-bosons.

To calculate the cross sections, we use the 
pQCD formalism, with the lowest-order hard
scattering cross sections found, for example, in
Refs.~\cite{ctqhbk,ehlq}.  For the production of jets, photons and heavy
quarks, we impose a cut $p_{T,\,{\rm min}} = 10 \, {\rm
GeV}$, which will keep us in the region where
perturbative calculations are applicable.  Since our main
interest is to probe parton densities at small $x$, we
will mostly need asymmetric configurations of the
momentum fractions, $x_{1}$ and $x_{2}$, of the partons
entering the hard scattering.  From the kinematic inequality
        $$
            x_{1} x_{2} \geq  \frac {4 p_{T,\,{\rm min}}^{2}}{s} ,
        $$
        we deduce that the momentum fractions obey $x_{1}x_{2} \geq
        2\times 10^{-6}$.
High-energy data on soft hadron production indicate \cite{UA5} that for fixed
$y_{max} -y$, where $y$ is the particle rapidity, the
soft hadronic multiplicity does not increase with energy, although at
$y=0$ it grows rapidly with $s$. Thus, for the large values of
$y$ that we use in this analysis, soft interactions result in much smaller
underlying event $E_T$ pedestals than for $y \sim 0$ at LHC energies. 
Moreover, for $x$ substantially larger than 
$x_{min}=2\times 10^{-6}/{\rm max}(x_1,x_2)$, 
the counting rates are so high for moderate $p_T$ of the jets (in processes
J$\gamma$ and JJ in expression (\ref{gnricpro}))
that it would be possible to restrict the analysis to the region of 
sufficiently large $x$ (for one of the partons)
so that the jets are
still produced at large $y$. Indeed, the expected
 rates are so high that it would be also
possible to check the role of the pedestals by taking data at several $x$
bins.
Hence, our choice of $p_{T,\,{\rm min}} = 10 \, {\rm GeV}$ appears safe.  

Since our interest is only in estimating rates and not in
a detailed extraction of parton densities from data,
it will be sufficient to perform leading-order
calculations for most of the processes considered.
Of course, when actual data become available,
it will be necessary to fit the parton densities to the
data with the aid of theoretical formulae at the best
possible accuracy, at least next-to-leading order.

At small $x$, the gluon density is substantially larger
than the quark densities.  Since at lowest order, the
Drell-Yan and vector-boson processes are given by quark-antiquark
annihilation, without a gluon-induced subprocess, we will
calculate these processes to next-to-leading order, with the
hard-scattering coefficients in the 
$\overline{\rm{MS}}$ scheme found in Ref.~\cite{vn}.

For jet, photon and heavy-quark production, we will set
the renormalization and factorization scale $\mu $ to the
commonly used value of $p_{T}$.  For Drell-Yan and
vector-boson production, we set $\mu$ to the
pair mass and vector-boson mass, respectively.

\section{Jet plus photon}
\label{sec:pj}

We calculate the cross section for producing a
jet and a photon, putting the events in bins of $p_{T}$ and $x$,
where $x$ is the minimum of the momentum fractions, $x_{1}$
and $x_{2}$, of the incoming partons.
The $x$ bin is defined by 
$x/\Delta\leq x \leq x\Delta$, where $\Delta=10^{1\over 2n_{div}}$
with $n_{div}=10$,
while the bin size in $p_T$ is set to 20\% of the 
central value up to $p_T \sim 110\ {\rm GeV}$ and 40\% beyond this
$p_T$ value. 
These same bin sizes are also used in the other cross sections
computed in this paper.

\begin{figure}
\centerline{
\psfig{file=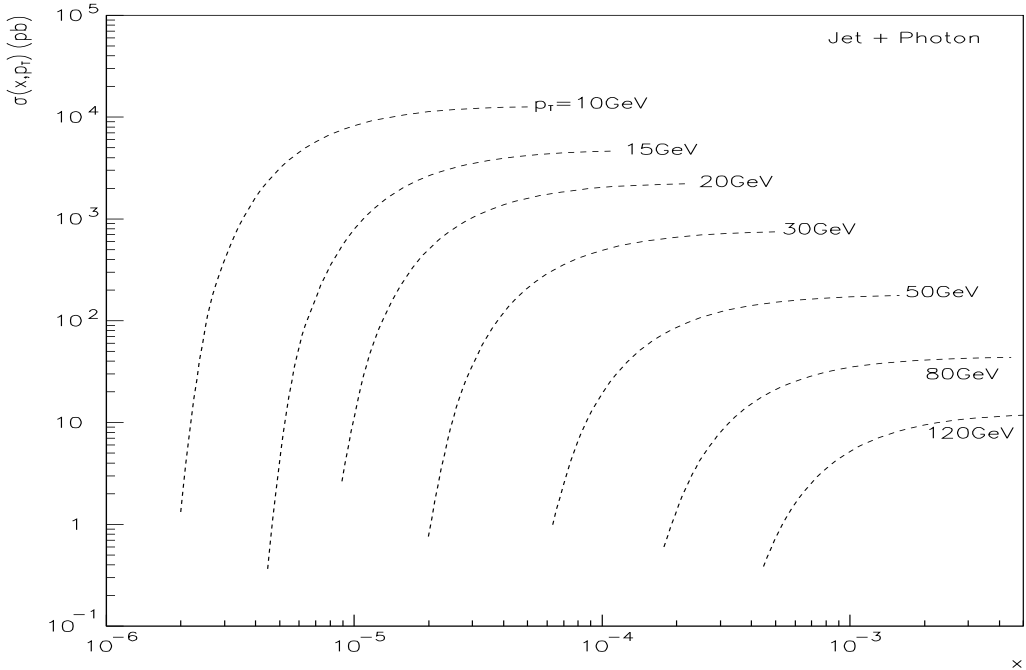,height=4in,width=6in,clip=}}
\caption{\sf The cross section $\sigma(x,p_T)$ for Jet $+ \ \gamma$ 
production as a function of $x={\rm min}(x_1,x_2)$ and $p_T$.  It is
integrated over a $p_T$ bin which is 20-40\% of the central value (see text),
max($x_1,x_2$)\ $<0.8$ and $x/\Delta<\ {\rm min}(x_1,x_2) < x\Delta$, where
$\Delta=10^{1/20}$.  This choice of $\Delta$ corresponds to 10 bins per decade
in $x$.}
\label{pjptx1}
\end{figure}

The cross sections shown in Fig.~\ref{pjptx1}
correspond to the kinematic region
\begin{equation}
(x_1\ll x_2\leq 0.8)\ \cup \ (x_2\ll x_1\leq 0.8).
\label{xregions}
\end{equation}
One sees that large cross sections (above 10 pb) are
obtained for $p_{T}$ up to about 100 GeV.  
Combined with a 
luminosity of $100 \, {\rm pb}^{-1}/{\rm yr}$, this gives at least
hundreds of events in every bin, which will give good
statistical precision.  The strong fall-off of the curves
at their left end is a result of approaching the
kinematic limit: $x > 4 p_{T,{\rm min}}^{2}/s$.  To get a
useful number of events, it is sufficient for $x$ to be
about twice its limit.

\begin{figure}
\centerline{
\psfig{file=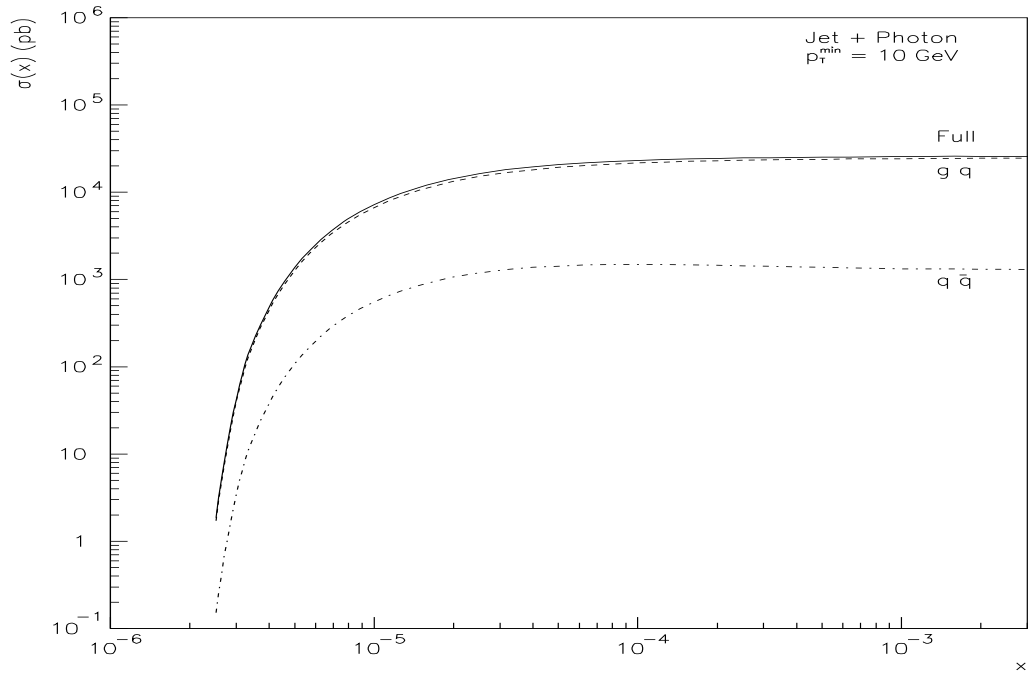,height=4in,width=6in,clip=}}
\caption{\sf The cross section $\sigma(x)$ for Jet $+ \ \gamma$ production as
a function of $x={\rm min}(x_1,x_2)$.  It is integrated over 
$p_T>10\ {\rm GeV}$ and the same region in $x_1,x_2$ as in the caption
for Fig.~\ref{pjptx1}.}
\label{pjx1}
\end{figure}

The cross section is dominated by gluon-quark scattering.
In Fig.~\ref{pjx1}, the cross section integrated over $p_{T} > 10 \,
{\rm GeV}$ is split into $gq$ and $q \bar q$ components,
and we see that the $gq$ term is about an order of
magnitude larger over the whole range shown. As before, the
fall-off at the left is a consequence of the
chosen minimum $p_{T}$.  Since the incoming quark is
typically at large $x$, where its distribution is already
known fairly accurately \cite{cteq3m}, 
photon-jet production provides a direct
measurement of the gluon density for small $x$ in the
range $x > 2.5 \times  10^{-6}$.

\begin{figure}
\centerline{
\psfig{file=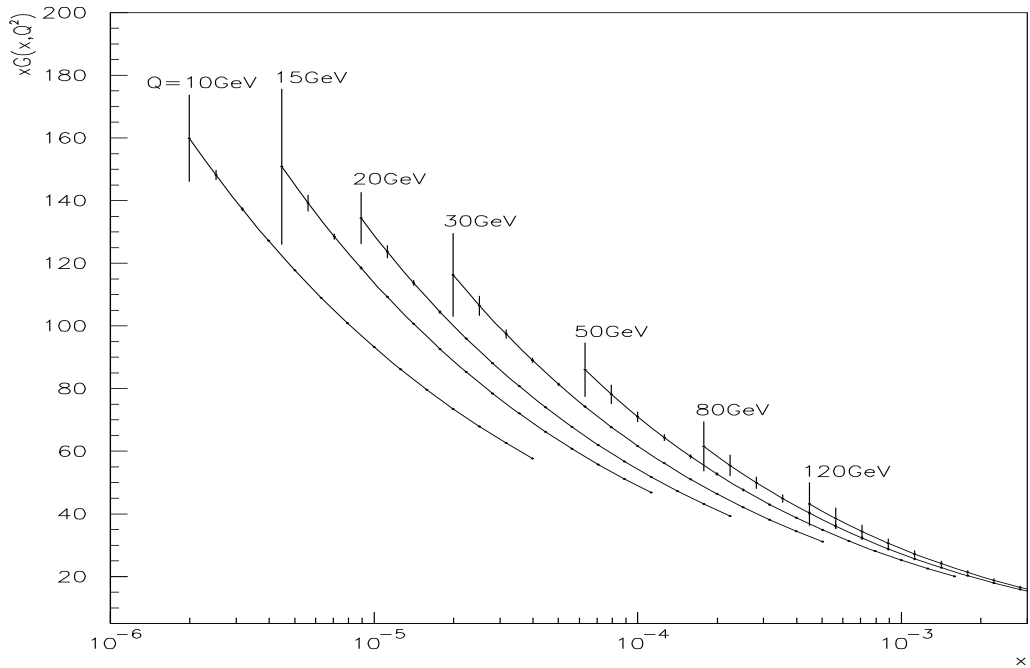,height=4in,width=6in,clip=}}
\caption{\sf The gluon momentum distribution with error bars
calculated from the Jet $+ \ \gamma$ cross section and a data sample
corresponding to an integrated luminosity of $100\ {\rm pb}^{-1}$.}
\label{xg}
\end{figure}

To illustrate the accuracy that can be achieved in a
determination of the gluon density, we show in Fig.~\ref{xg} the
CTEQ3M gluon momentum density $xG(x,Q^{2}=p_T^2)$ together with
the statistical errors on measurements that correspond to
the cross sections in Fig.~\ref{pjptx1}.  Notice that there is
sufficient precision not merely to measure the gluon
density but also to test its evolution.
The cross sections presented here are, of course, the result of
an extrapolation of parton densities from a region where they have been
measured to much smaller $x$ values.

\section{Lepton Pairs}
\label{sec:dy}

\begin{figure}
\centerline{
\psfig{file=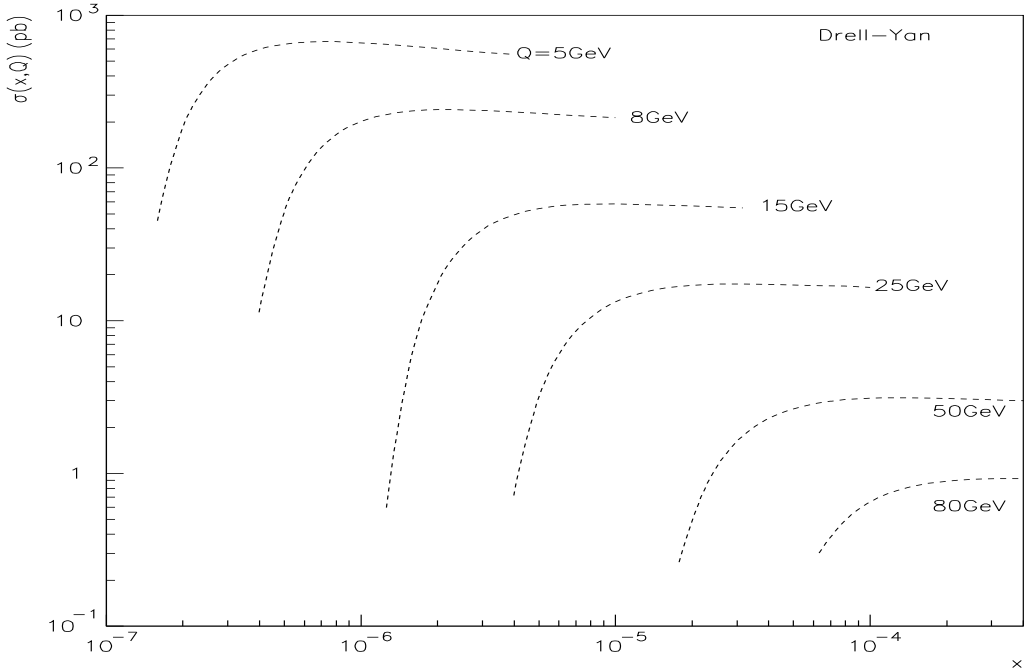,height=4in,width=6in,clip=}}
\caption{\sf The cross section $\sigma(x,Q)$ for Drell-Yan
as a function of $x={\rm min}(x_1,x_2)$ and $Q$.  It is
integrated over a $Q$ bin which is 20-40\% of the central value
and the same region in $x_1,x_2$ as in the caption
for Fig.~\ref{pjptx1}.}
\label{dyxM}
\end{figure}

We next consider lepton pair production from hadrons (the Drell-Yan process).
Using the same kinematic region defined by expression (\ref{xregions}), 
we obtain the next-to-leading order cross sections shown in
Fig.~\ref{dyxM}.  
Again, the relatively large cross sections lead to a large number of
events expected at the LHC.  
More significantly, by measuring cross sections at smaller values of the 
pair mass $Q$, we can obtain parton densities at lower values of $x$
than in the other processes considered
in this paper.  The advantage of the Drell-Yan process is that one can
go to fairly small $Q$ values while still trusting the pQCD formalism. 

\begin{figure}
\centerline{
\psfig{file=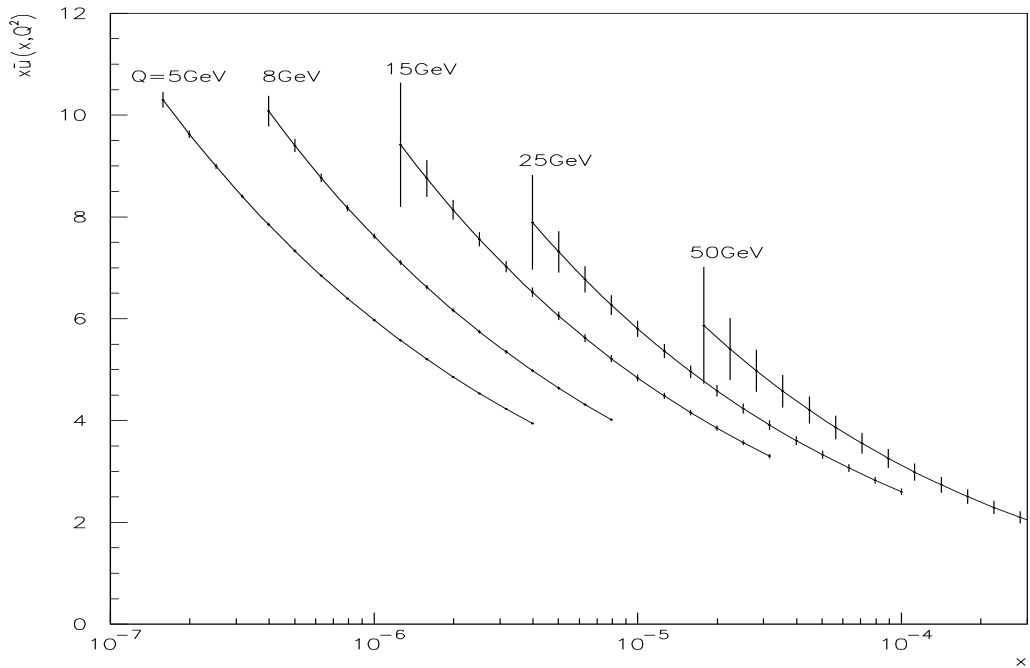,height=4in,width=6in,clip=}}
\caption{\sf The ${\bar u}$ antiquark momentum distribution with error bars
calculated from the Drell-Yan cross section and a data sample
corresponding to an integrated luminosity of $100\ {\rm pb}^{-1}$.}
\label{uq}
\end{figure}

The dominant contribution for this process comes from the $u{\bar u}$
channel; the $qg$ contributions are about 30 times smaller than 
those from $q{\bar q}$.
Using quark distributions $q(x_2)$ which are well-determined
for large $x_2$, one may then extract antiquark densities ${\bar q}(x)$
at small $x$.  We show in Fig.~\ref{uq} a plot of the extrapolated CTEQ3M
antiquark density $x{\bar u}(x,Q^2)$ with the statistical errors based on 
the cross sections shown in Fig.~\ref{dyxM}.
Note that we consider the Drell-Yan process down to relatively small pair
masses 
$\sim 5\ \rm{GeV}$.  This assumes that, in the experiment, the 
detector will be able to suppress the background due to heavy flavor 
decays.  Such suppression might be achieved using information from the
forward calorimeters as well as from microvertex detectors.

\section{Two Jets}
\label{sec:twoj}

\begin{figure}
\centerline{
\psfig{file=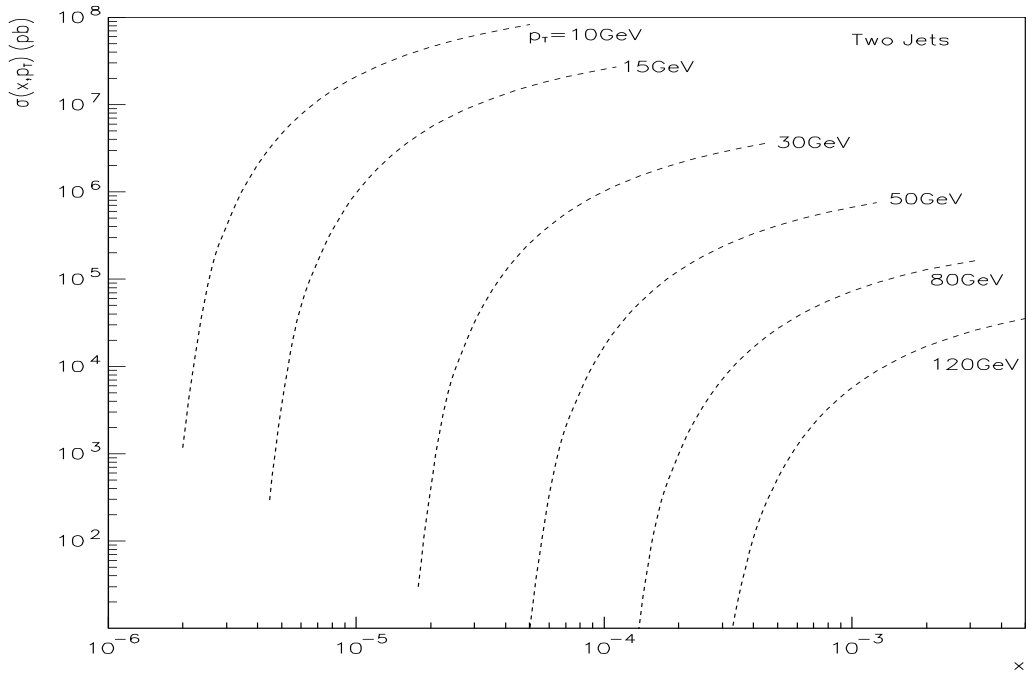,height=4in,width=6in,clip=}}
\caption{\sf The cross section $\sigma(x,p_T)$ for dijet production as a
function of $x={\rm min}(x_1,x_2)$ and $p_T$.
Same comments as in the caption for Fig.~\ref{pjptx1} apply.}
\label{jjptx1}
\end{figure}

We next consider the cross section for the production of dijets.
Fig.~\ref{jjptx1} shows the cross section in 
bins of $p_T$ and $x$, in the region defined by expression (\ref{xregions}).
The cross sections obtained are about 100 times larger than all the others
computed in this paper.  

\begin{figure}
\centerline{
\psfig{file=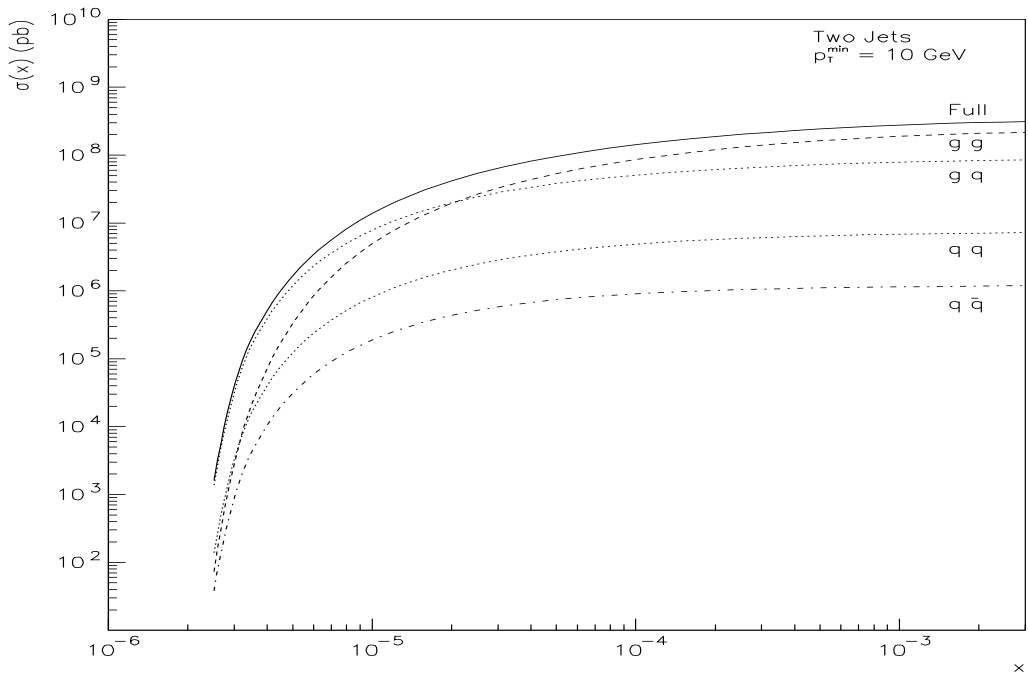,height=4in,width=6in,clip=}}
\caption{\sf The cross section $\sigma(x)$ for dijet production as
a function of $x={\rm min}(x_1,x_2)$.  
Same comments as in the caption for Fig.~\ref{pjx1} apply.}
\label{jjx1}
\end{figure}

In Fig.~\ref{jjx1}, we show the $x$ distribution of the cross section.
Aside from the total cross section (solid curve), we also
exhibit the contributions from the different partonic channels:
$gg$ (dashed), $gq$ (upper dotted), $qq$ (lower dotted) 
and $q{\bar q}$ (dot-dashed curve).  
The largest contributions to the cross section come from
the $gg$ and $gq$ channels.
This process thus provides an independent consistency check of the
parton distributions, primarily the gluon density,
obtained from other processes.

\section{$W$ Production}
\label{sec:wpro}

Production of $W$-bosons in the forward direction
provides a complementary way to measure sea quark
distributions at small $x$.  At the same time, comparison
of the rates of production of $W^{+}$ and $W^{-}$ bosons would
provide a new method to measure the ratio $r(x,Q^{2}) =
\frac {u(x,Q^{2})}{d(x,Q^{2})}$ at large $x$ \cite{W}.  (Such a 
measurement would
rely on the presumed equality of $\bar u$ and $\bar d$
distributions at small $x$.) Current determinations of
$u/d$ rely on data from deuterium targets, where nuclear
effects can be large for $x\geq 0.5$. Notice that most of the
current global fits, eg.~\cite{cteq3m}, 
assume that $r \to  0$ when $x\to 1$, whereas a
perturbative QCD analysis of the leading diagrams for the
$x\to 1$ limit \cite{FJ} suggests that $r\to 0.2$ in this limit.
A direct measurement independent of any data requiring
knowledge of nuclear effects would be valuable.

\begin{figure}
\centerline{
\psfig{file=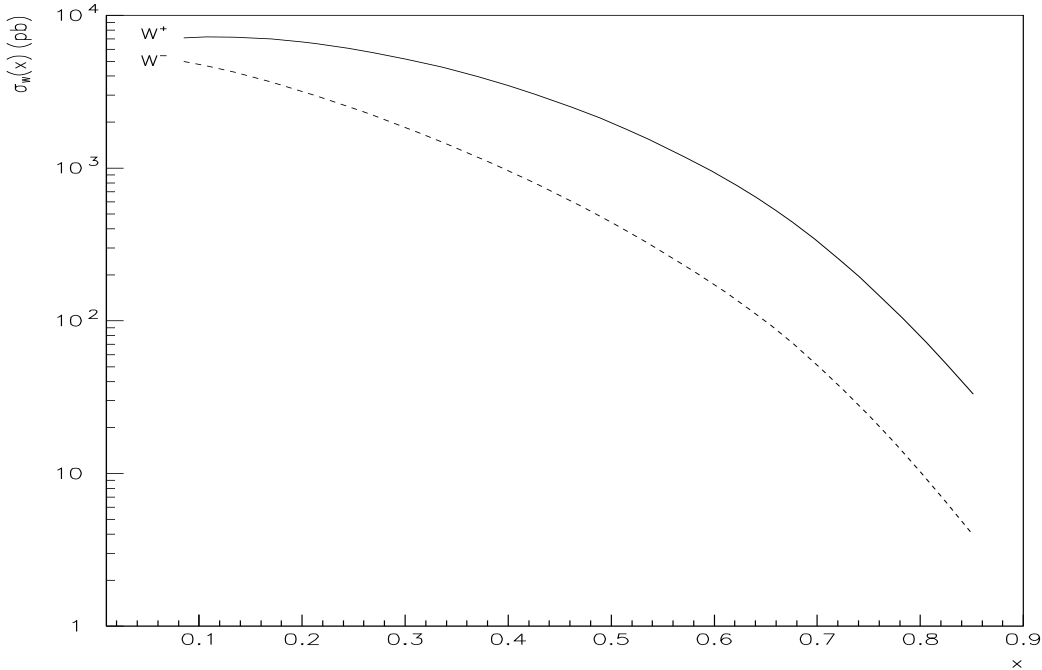,height=4in,width=6in,clip=}}
\caption{\sf The cross section $\sigma(x)$ for $W$ production as a
function of $x=x_1$.  It is integrated over 
$x/\Delta< x < x\Delta$, where $\Delta=10^{1/20}$.  
This choice of $\Delta$ corresponds to 10 bins per decade
in $x$.  Note that for this process, $x_2=M_W^2/(xs)$.}
\label{wvx}
\end{figure}

At next-to-leading order, we calculate the cross sections for $W^+$ and
$W^-$ production in bins of $x$. 
One observes from Fig.~\ref{wvx} that significant 
rates can be measured up to $x \sim 0.8$. This implies that
quark distributions at $Q^2 \sim 10^4\ {\rm GeV}^2$ can be measured from 
the sum of the cross sections for production
of $W^+$ and $W^-$-bosons down to $x \sim 5\times 10^{-5}$. At the same 
time, one would be able to distinguish between 
different scenarios for the asymptotic behaviour of the $u/d$ ratio in 
the $x \rightarrow 1 $ limit.

\section{Heavy Quarks}
\label{sec:hq}

\begin{figure}
\centerline{
\psfig{file=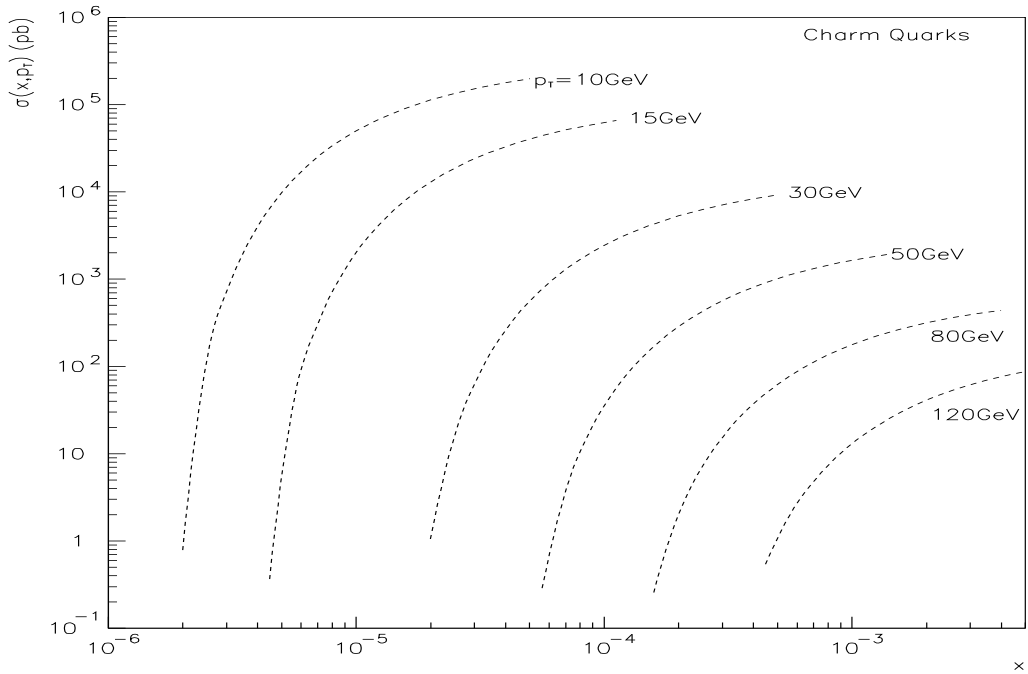,height=4in,width=6in,clip=}}
\caption{\sf The cross section $\sigma(x,p_T)$ for charm quark production
as a function of $x={\rm min}(x_1,x_2)$ and $p_T$.  Same comments as in
the caption for Fig.~\ref{pjptx1} apply.}
\label{cqptx1}
\end{figure}

\begin{figure}
\centerline{
\psfig{file=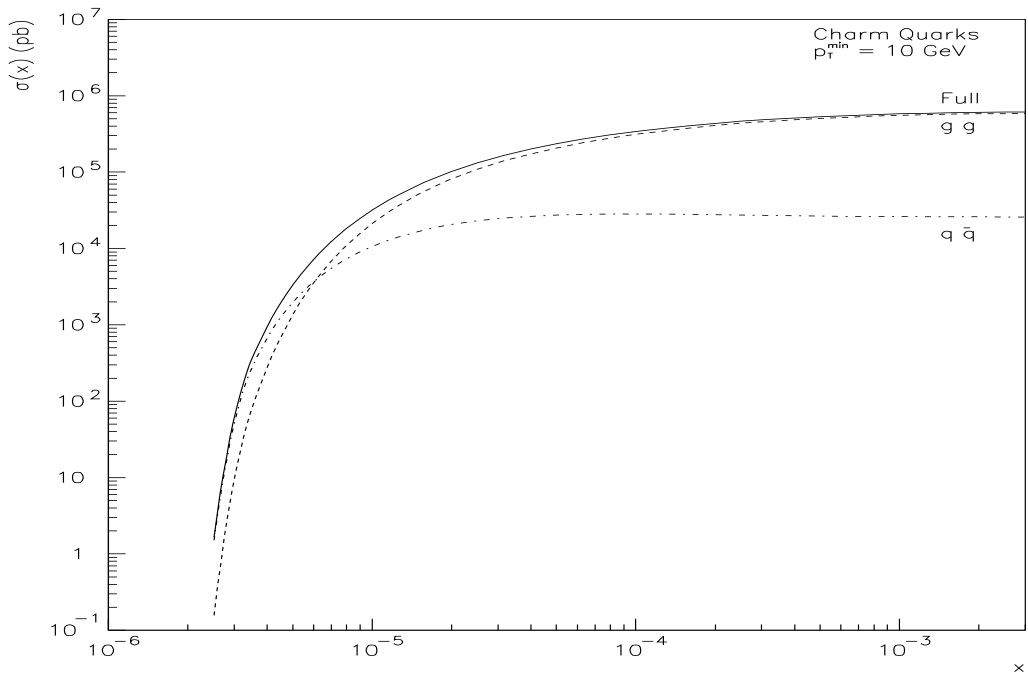,height=4in,width=6in,clip=}}
\caption {\sf The cross section $\sigma(x)$ for charm quark production as
a function of $x={\rm min}(x_1,x_2)$.  
Same comments as in the caption for Fig.~\ref{pjx1} apply.}
\label{cqx1}
\end{figure}

We show in Figs.~\ref{cqptx1} and \ref{cqx1} the $(x,p_T)$ 
and $x$ distributions, respectively, for charm quark production
in the region defined by expression (\ref{xregions}).
The same general features as in the plots for 
jet plus photon and dijets are observed.


In Fig.~\ref{cqx1}, the contributions from the two active
partonic channels are also shown:
$gg$ (dashed curve) and $q{\bar q}$ (dot-dashed curve).  
The total cross section is of order 1 to $10^2$ pb when
$2.5\times 10^{-6} \leq x \leq 4\times 10^{-6}$.  In this region, the
$q{\bar q}$ contribution dominates, being about 10 times larger than that
of the $gg$ channel.
In the region $10^{-4} \leq x \leq 10^{-3}$,
the total cross section rises, is of order $10^5$ pb and mostly consists
of $gg$ channel contributions; the $q{\bar q}$ contribution drops to 
about 10\% of the total.

Thus, it is possible to extract the antiquark (or quark) density 
at $x \sim 3-4\times 10^{-6}$ from the charm quark cross section
providing a cross-check of the antiquark densities obtained from
Drell-Yan measurements.
In this case, the background
consisting of $gg$ channel contributions is well determined with the
use of the gluon density obtained from the jet plus photon or dijet
cross section as described in Sections \ref{sec:pj} and \ref{sec:twoj}.
Similarly, it is also possible to extract the gluon density at
$x \geq 10^{-4}$ from this type of cross section, providing
a cross-check of gluon densities derived from other measurements.

\section{Conclusions}
\label{sec:conc}

\begin{figure}
\centerline{
\psfig{file=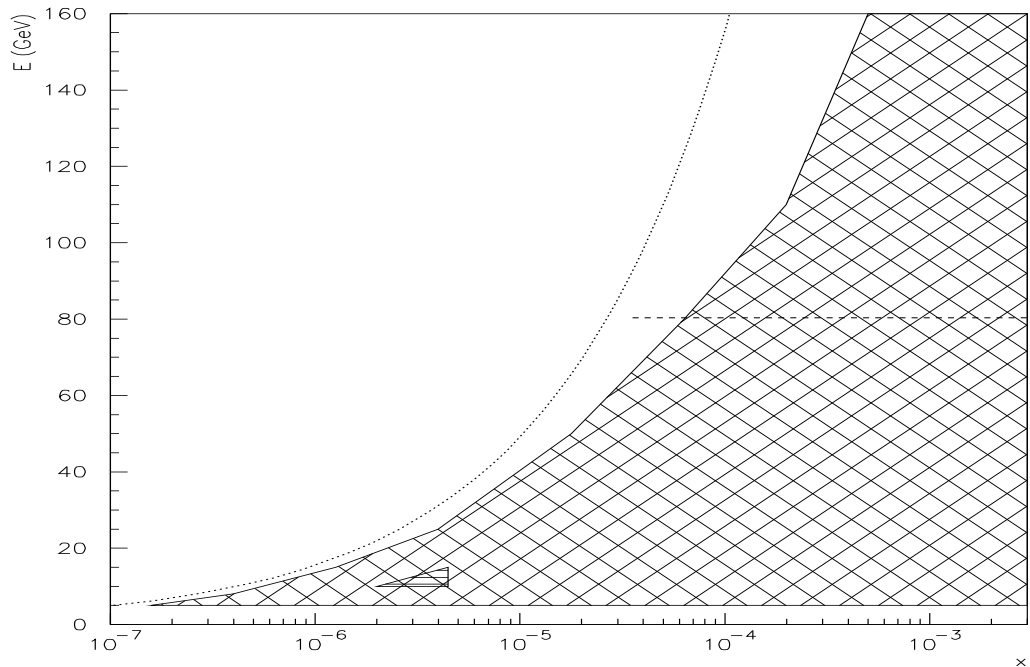,height=4in,width=6in,clip=}}
\caption{\sf A contour plot showing the region in $(E,x)$ where
measurements of the antiquark densities in the proton can be made to
within 20\% accuracy, assuming an integrated luminosity 
of $100\ {\rm pb}^{-1}$.  
The region marked with cross-hatch shading corresponds
to the Drell-Yan process.  The dashed horizontal line is where $W$
production dominates.  The dotted curve shows the minimum value of $x$
as a function of the pair mass in the Drell-Yan process. 
The small triangle with horizontal shading shows where
charm production is dominated by the $q{\bar q}\rightarrow c{\bar c}$
subprocess.}
\label{ptxqctr}
\end{figure}

\begin{figure}
\centerline{
\psfig{file=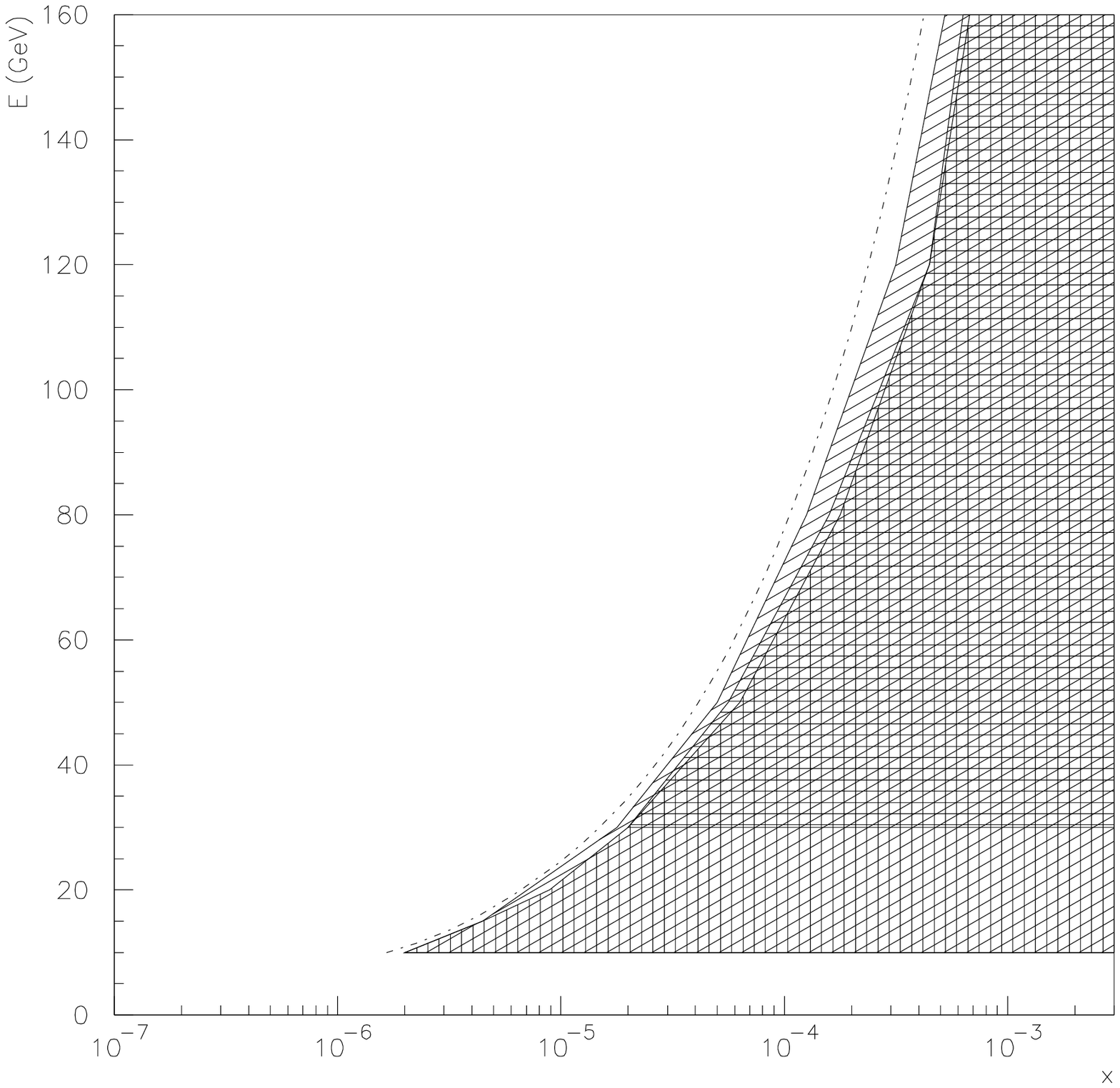,height=4in,width=6in,clip=}}
\caption{\sf A contour plot showing the region $(E,x)$ where measurements 
of the gluon density in the proton can be made to within 20\% accuracy, 
assuming an integrated luminosity of $100\ {\rm pb}^{-1}$.
The dot-dashed curve represents the minimum allowed value of
$x$ as a function of the $p_T$ of the jet or heavy quark.
The regions marked with vertical, horizontal and slanted lines correspond 
to jet plus $\gamma$, charm quark and dijet production, respectively.}
\label{ptxGctr}
\end{figure}

We have shown, in several independent ways, how parton distributions in 
nucleons and nuclei (nuclear shadowing) down to $x\sim 10^{-7}$
may be measured from hadron-hadron interactions at LHC energies.
In particular, we show in Figs.~\ref{ptxqctr} and \ref{ptxGctr}
the regions in energy $E=(p_T, Q\ {\rm or}\ M_W)$
and momentum fraction $x$ where
measurements can be made for extracting 
antiquark and gluon densities, respectively, at
better than 20\% accuracy from the different hadronic processes
considered, assuming an integrated luminosity of $100\ {\rm pb}^{-1}$.

In Fig.~\ref{ptxqctr}, the region marked with cross-hatch
shading shows where the Drell-Yan process determines the
antiquark densities with better than 20\% accuracy.  The
dashed horizontal line is where $W$ production determines
these densities.  The small triangle with horizontal shading
shows where charm production determines the antiquark
densities.  The dotted curve shows the minimum value of $x$
as a function of the pair mass $Q$
in the Drell-Yan process.

Similarly, we show in Fig.~\ref{ptxGctr} the regions where data on
the different processes can determine the gluon density.
The regions marked with vertical,
horizontal and slanted lines correspond to 
jet plus photon, charm quark and dijet production, respectively.
The dot-dashed curve represents the minimum allowed value of
$x$ as a function of the $p_T$ 
of the jet or heavy quark.

In our calculations, we have assumed that the usual DGLAP
evolution equations apply to the parton densities in the very
small $x$ region being considered and that CTEQ3M parton distributions
can be extrapolated outside the region in which they are valid.
However, it may well be
that other physics effects, notably gluon recombination \cite{gluerec},
are important at these small $x$ values.
As we have seen, by measuring 
parton densities over a
wide range of scales and in several processes, one
can, in fact, test whether DGLAP evolution and the rest of the
factorization method is valid --- see, for example, Figs.~\ref{xg}
and \ref{uq}.  Of course, if the measurements fail consistency
checks, for example if DGLAP evolution is found to fail, then
one can no longer say that it is the parton densities that
are being measured, but rather that one is probing new
effects at small $x$.

\section*{Acknowledgments}

This work was supported in part by the U.S.~Department of Energy under
grant numbers DE-FG02-90ER-40577 and DE-FG02-93ER40771 
and by the U.S.~National Science Foundation.

\end{document}